\documentclass{article}
\usepackage {graphicx}
\begin{document}
\baselineskip .75cm 
\begin{titlepage}
\title{\bf Self-consistent quasiparticle model for quark-gluon plasma}       
\author{Vishnu M. Bannur  \\
{\it Department of Physics}, \\  
{\it University of Calicut, Kerala-673 635, India.} }   
\maketitle
\begin{abstract}

Here we present a self-consistent quasi-particle model for quark-gluon 
plasma and apply it to explain the non-ideal behaviour seen in 
lattice simulations. The basic idea, borrowed from electrodynamic plasma, 
is that the gluons acquire mass as it propagates through plasma due to 
collective effects and is approximately equal to the plasma frequency. 
The statistical mechanics and thermodynamics of such a system is studied 
by treating it as an ideal gas of massive gluons. Since mass or plasma 
frequency depends on density, which itself is a thermodynamic quantity, 
the whole problem need to be solved self-consistently.    
\end{abstract}
\vspace{1cm}
                                                                                
\noindent
{\bf PACS Nos :} 12.38.Mh, 12.38.Gc, 05.70.Ce, 52.25.Kn \\
{\bf Keywords :} Equation of state, quark-gluon plasma, 
quasiparticle quark-gluon plasma. 
\end{titlepage}
\section{Introduction :}

The quasiparticle model of quark-gluon plasma (qQGP) is used to study the 
thermal properties of quark-gluon plasma (QGP) \cite{pe.1,pe.2,lh.1,s.1,ba.1}. 
QGP is a plasma of 
quarks and gluons which exhibits a collective behaviour because of the 
interactions,  governed by quantum chromodynamics (QCD). At extreamly high 
temperature, the interaction may be very weak because of the asymptotic 
freedom and hence the plasma is almost ideal gas of quarks and gluons. 
However, near and above the critical $T_c$, the coupling constant 
$\alpha_s$ is not weak and hence QGP may be non-ideal gas of quarks 
and gluons. Here $T_c$ is the critical temperature of phase transition 
or cross over from hadrons to QGP. 
The lattice simulation of QCD at finite temperature \cite{ka.1} 
(LGT) and the elliptical flow observed in relativistic heavy ion 
collisions (RHICs) signal the existance of non-ideal QGP. The problem of 
confinement-deconfinement of QCD and the search for QGP in RHICs 
depend on the properties of QGP at temperature close to $T_c$. 
Since the strong coupling constant $\alpha_s$ is not very small 
near $T_c$, it may be difficult to solve QCD analytically. Many  
attempts in this line were not completely successful \cite{bl.1,d.1,kj.1}. 
Therefore, several phenomenological models with two or more fitting parameters, 
like confinement models \cite{ba.2}, strongly coupled plasma (SCQGP) 
\cite{ba.3}, strongly interacting plasma (sQGP) \cite{sh.1}, 
qQGP \cite{pe.1,pe.2,lh.1,s.1,ba.1}, liquid model \cite{ra.1} 
and so on, were proposed. 
By adjusting parameters of the model one may fit the LGT results.  
Here we propose a new quasiparticle model which has only one adjustable 
parameter and is the revised version of the widely studied qQGP 
\cite{ba.1,pe.1,go.1}.      

\section{Self-consistent qQGP:} 

Let us recollect first various versions of qQGP models. It was first 
proposed by Peshier {\it et. al.} \cite{pe.1} and with two adjustable 
parameters, they were were able to fit LGT results of \cite{l1.1}   
on gluon plasma. However, it fails to fit the more recent LGT 
results as shown in \cite{ba.3}. Furthermore, Gorenstein and Yang \cite{go.1} 
pointed out that there is a thermodynamic (TD) inconsistency in this model 
and by imposing a stringent constraints, which they called TD consistency 
relation or the reformulation of statistical mechanics (SM), they tried 
to solve the problem. Recently, in a 
series of papers \cite{ba.1}, we pointed out the reason for the TD 
inconsistency and revised their model using the standard SM in a TD 
consistent way. The revised model of qQGP, without TD consistency relation 
or the reformulation of SM, fits remarkable well many LGT results with 
a single system dependent parameter \cite{ba.1}. We also showed how the 
TD consistency relation comes out from our formulation. The basic idea 
in our model is that we start from the energy density of quasiparticles 
instead of ideal gas pressure-partition function relation. Further, we 
assume that quasiparticle acquire an additional temperature dependent 
mass and this thermal mass is equal to the plasma frequency, $\omega_p$, 
because of the collective effects of QGP. We used, as an approximation, the 
expression for $\omega_p$ from perturbative QCD results and assumed the   
two-loop running coupling constant as function of temperature. 
So we just have one adjustable parameter related to QCD scale parameter 
which appears in $\alpha_s(T)$. In this letter, we further revise 
our qQGP model by considering a density dependent plasma frequency, 
instead of the perturbative QCD results which is valid at extreamly high 
temperature. It is motivated from a similar work in ultra-relativistic 
($e$, $e^+$ and $\gamma$) system \cite{me.1,ba.4} where one uses 
ultra-relativistic plasma frequency which depends on density 
and temperature.       

Here we consider a simplest and well studied system, gluon plasma (GV). 
It is a system of gluons interacting via QCD interaction at finite 
temperature, $T$. As we discussed earlier, it is difficult to solve 
it analytically since the coupling constant is not weak. Therefore, 
we consider a phenomenological model where the thermal properties may be 
obtained by studying the thermal excitations of plasma modes. These 
thermal excitations are called quasi-gluons with the quantum numbers of 
gluons and with the thermal mass equal to the plasma frequency. 
Thus we have a gas of non-interacting or ideal quasi-gluons and 
following the standard SM \cite{pa.1}, the logaritham of grand 
partition function or q-potential is given by 
\begin{equation} 
  q = - \sum_{k=0}^\infty \ln (1 -  
e^{ - \beta \epsilon_k})\,\, , \label{eq:p}  
\end{equation}  
where $\epsilon_k$ is the single particle energy of quasi-gluon, 
i.e, gluon with temperature dependent mass, given by,
\[ \epsilon_k = \sqrt{k^2 + m^2 (T)} \,\,, \]   
where $k$ is the momentum and $m$ is the thermal mass which is equal to 
the plasma frequency. $\beta$ is defined as $1/T$. 
Instead of using the QCD perturbative expression for plasma frequency 
as done in the earlier qQGP models \cite{pe.2,ba.1}, we model it as 
\begin{equation}
\omega_p^2 \propto \alpha_s \, \frac{n}{T} \equiv a_0\,\alpha_s \, 
\frac{n}{T}\,\,, 
\end{equation}
which is motivated from a similar work on electrodynamic plasma (EDP) 
\cite{me.1,ba.4}. The constant $a_0 = \frac{8\pi}{3}$ in electron-positron 
EDP plasma and here we fix it by demanding that 
$\omega_p^2 \rightarrow \frac{g^2 \,T^2}{3}$ 
as $T \rightarrow \infty$, the QCD perturbative result. 
$g^2$ is related to $\alpha_s$ through the relation $\alpha_s = g^2/ 4 \pi$.   
Further, we use temperature dependent running coupling constant, 
$\alpha_s (T)$, which is motivated from lattice simulation of QCD \cite{ka.1}.  
$n$ is the density of gluons which is taken 
to be the same as the number of quasi-gluons in quasiparticle models. 
Again from standard SM, 
\begin{equation}
n =  \frac{g_f}{2 \pi^2}\,\int_0^{\infty} dk\,k^2\, \frac{1}{
 e^{\beta \sqrt{k^2 + a_0\,\alpha_s\,\frac{n}{T}}} - 1} \,\,, 
\label{eq:n0} \end{equation}
which may be rewritten as 
\begin{equation}
n = \frac{g_f}{2 \pi^2}\,T^3\,\int_0^{\infty} dx\,x^2\, \frac{1}{
 e^{ \sqrt{x^2 + a_0\,\alpha_s\,\frac{n}{T^3}}} - 1} \,\,, 
\label{eq:n1} \end{equation} 
where $g_f = 16$, the degeneracy associated with the internal 
degrees of freedom. These equation need to be solved 
self-consistently because $n$ which is to be determined is inside 
the integral through $\omega_p$. Note that we don't use the perturbative 
QCD expression for plasma frequency, which is appropriate at 
$T \rightarrow \infty$, instead we calculate it self-consistently. 
Redefining the variables, 
the final equation to be solved self-consistently is,  
\begin{equation}
f_g^2 = \int_0^{\infty} dx\,x^2\, \frac{1}{
 e^{\sqrt{x^2 + \bar{a}^2\,f_g^2 }} - 1} \,
= \bar{a}^2 \,f_g^2 \,\sum_{l=1}^{\infty} \,\frac{1}{l}\,
K_2 (\bar{a}\,l\,f_g) \,\,,   
\label{eq:fg0} \end{equation} 
where 
\[ \bar{a}^2 \equiv \frac{g_f}{2\,\pi^2} \,a_0\,\alpha_s \,,\] 
and 
\[f_g^2 = \frac{2 \pi^2}{g_f}\,\frac{n}{T^3}\,\,,\]    
where $K_2$ is the modified Bessel function. 
Above equation, Eq. (\ref{eq:fg0}), may be solved numerically 
to get $f_g^2$ and then other TD quantities like energy density, 
pressure etc. may be calculated. 

The energy density is given by,
\begin{equation}
\varepsilon = \frac{g_f}{2 \pi^2}\,T^4\,
\int_0^{\infty} dx\,x^2\, \frac{\sqrt{x^2 + \bar{a}^2 \, f_g^2 }}{
 e^{\sqrt{x^2 + \bar{a}^2 \, f_g^2 }} - 1} \,\,, 
 \end{equation}
or 
\begin{equation}
\varepsilon = \frac{g_f}{2\,\pi^2}\,T^4 \sum_{l=1}^{\infty} 
\left. \left[ (\bar{a} f_g l)^3 
K_1 (\bar{a} f_g l) +  3\, (\bar{a} f_g l)^2 K_2 (\bar{a} f_g l) 
\right] \right)\,\,, \label{eq:ef} 
\end{equation}
in terms of modified Bessel functions $K$. Next, the pressure may be 
obtained from the TD relation, 
\begin{equation}
\varepsilon =  T \frac{\partial P}{\partial T} 
- P \,\, ,  \label{eq:td} 
\end{equation}
on integration, which is the procedure we follow here. In this analysis, 
we have neglected vacuum energy or zero point energy  
with the assumption that the whole thermal properties of 
gluon plasma may described by the quasi-gluon excitations. 
A new TD consistent qQGP model with the vacuum energy is presented in 
Ref. \cite{ba.1}. 

\begin{figure}[h]
\centering
\includegraphics[height=8cm,width=12cm]{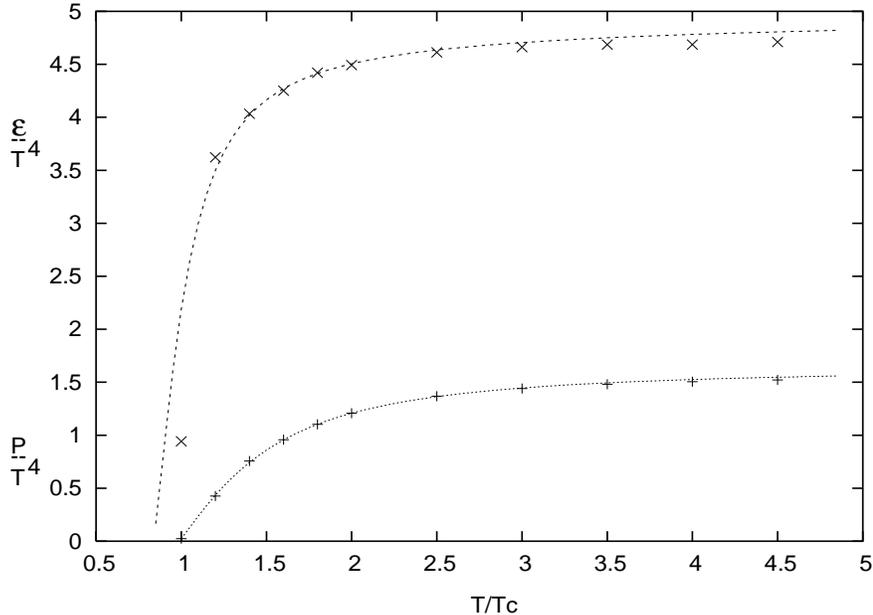}
\caption { Plots of $P/ T^4 $ (lower set of graphs) and  
$\varepsilon/ T^4$ (upper set of graphs)  as a function of $T/T_c$ from 
our model and lattice results (symbols) for gluon plasma.} 
\end{figure}

\section{Results:} 

In Fig. 1, we plotted energy density and pressure 
of GP from our model along with the LGT results \cite{ka.1}. First we 
solve, self-consistently, the integral equation for the density, 
Eq. (\ref{eq:fg0}), for a given temperature and obtain $f_g (T)$. We have used 
2-loop order running coupling constant, $\alpha_s(T)$, which is similar 
to one used in LGT calculations, and is given by, 
\begin{equation} \alpha_s (T) = \frac{6 \pi}{(33-2 n_f) \ln (T/\Lambda_T)}
\left( 1 - \frac{3 (153 - 19 n_f)}{(33 - 2 n_f)^2}
\frac{\ln (2 \ln (T/\Lambda_T))}{\ln (T/\Lambda_T)}
\right)  \label{eq:ls} \;, \end{equation}
where $\Lambda_T$ is a parameter related to QCD scale parameter and 
$n_f$ is the number of flavors which is zero in our case. 
Once we know $f_g (T)$, energy density and pressure are calculated using 
Eq. (\ref{eq:ef}) and Eq. (\ref{eq:td}) respectively. We adjust only one 
parameter of our model, $t_0 \equiv \Lambda_T/T_c$, so that we get 
the best fit to the energy density of LGT results \cite{ka.1}. 
We found that $t_0 = 0.83$. We need one more integration constant, which is 
not a parameter, to evaluate pressure which we fix to LGT data at $T = T_c$.  
We may as well fix it at $T = \infty$ such that $P/P_{SB} = 1$, 
where $P_{SB}$ is the ideal gas pressure, but it is difficult to solve 
numerically.   

\begin{figure}[h]
\centering
\includegraphics[height=8cm,width=12cm]{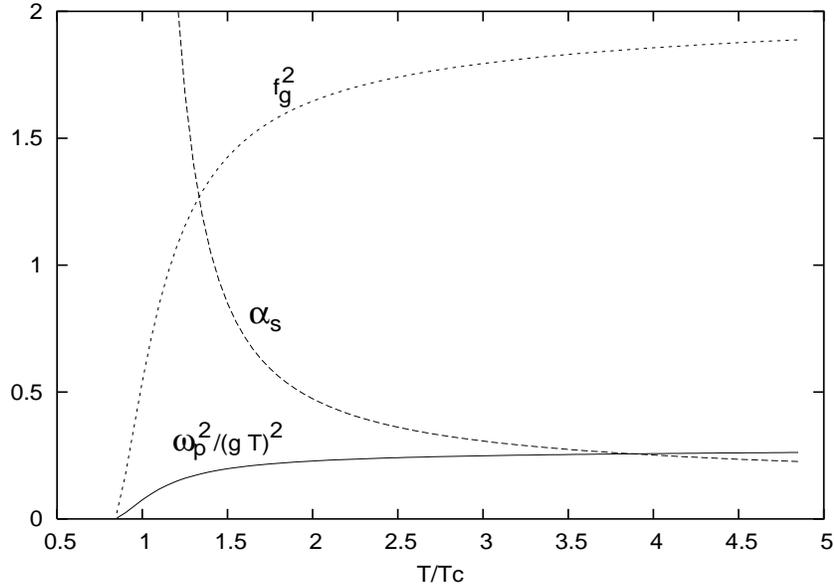}
\caption { Plots of $f_g^2 (T)$, $\alpha_s (T)$ and $\omega_p^2 / 
(g^2\,T^2)$ as a function of $T/T_c$.} 
\end{figure}

In Fig. 2, we plotted $f_g^2 (T)$, $\alpha_s (T)$ and $\omega_p^2 / 
(g^2\,T^2)$ as a function of $T/T_c$. The running coupling 
constant $\alpha_s (T)$ increases rapidly as $T \rightarrow T_c$ and 
$\omega_p^2 / (g^2\, T^2)$ is small near $T=T_c$ and 
asymptotically approaches $1/3$, the QCD perturbative result.   

\section{Conclusions:} 

We presented a new quasiparticle model for gluon plasma where gluons 
acquire mass, approximately equal to the plasma frequency, due to 
collective effects. Since $\omega_p$ depends on density, which is 
one of the thermodynamic quantity to be determined from statistical 
mechanics, we solved the problem self-consistently. Using this 
result, energy density and pressure were evaluated and by adjusting 
a single parameter, related to QCD scale parameter, we got a remarkable 
good fits to LGT results. Further extension of the model to QGP 
with quarks may be interesting, but not as simple as GP.

\end{document}